\documentclass[fleqn,usenatbib]{mnras}
\usepackage[english]{babel}

\usepackage[T1]{fontenc}
\usepackage{ae,aecompl}

\usepackage{fancyhdr}
\usepackage{amsfonts}
\usepackage{amsmath}
\usepackage{amssymb}
\usepackage{multicol}
\usepackage{layout}
\usepackage{graphicx}
\usepackage{multirow}

\usepackage{color}
\usepackage{multirow}
\usepackage[utf8]{inputenc}
\usepackage{times}
\usepackage{natbib}

\def\aap{A\&A}
\def\apj{ApJ}

\def\apjl{ApJ}
\def\mnras{MNRAS}

\def\nat{Nature}

\def\prd{Phys. Rev. D}

\title[Gravitational Grating]{Gravitational Grating}

\author[Sohrab Rahvar ]{Sohrab Rahvar$^{ }$\thanks{E-mail: rahvar@sharif.edu}\\
$^{ }$Physics Department, Sharif University of Technology, P.O.Box 11155-9161, Azadi Avenue, Tehran, Iran}

\graphicspath{{./fig/}}

\begin{document}






	
 \maketitle

\begin{abstract}
In this work, we study the interaction of the electromagnetic wave (EW) from a distant quasar
with the gravitational wave (GW) sourced by the binary stars.  While in the regime of geometric optics, the light bending due to this interaction is
negligible, we show that the phase shifting on the wavefront of an EW can produce the diffraction pattern on the observer plane. The diffraction of the light (with the wavelength of $\lambda_e$) by the gravitational wave playing the role of {\it gravitational grating} (with the wavelength of $\lambda_g$) has the diffraction angle of $\Delta\beta \sim \lambda_e/\lambda_g$.
The relative motion of the observer, the source of gravitational wave and the quasar results in a relative motion of the observer through the interference pattern on the observer plane. The consequence of this fringe crossing is the modulation in the light curve of a quasar with the period of few hours in the microwave wavelength. 
The optical depth for the observation of this phenomenon for a Quasar with the multiple images strongly lensed by a galaxy where the light trajectory of some of the images crosses the lensing galaxy is $\tau \simeq 0.2$. By shifting the time-delay of the light curves of the multiple images in a strong lensed quasar and removing the intrinsic variations of a quasar, our desired signals, as a new method for detection of GWs can be detected.

\end{abstract}

\begin{keywords}
gravitational waves; gravitational lensing; waves
\end{keywords}

\section{Introduction}
\label{introduction}

The first direct detection of gravitational wave is done by the Advanced LIGO experiment where they could detect the merging of the two blackholes in a binary system \citep{LIGO2016}. While the gravitational wave has already been detected indirectly by studying the orbital period of neutron stars \citep{taylor}, direct detections of gravitational waves opened a new window in the astronomy besides the two windows of the electromagnetic and the high energy cosmic ray particles. The observation of gravitational wave by LIGO and VIRGO experiments \citep{LIGO2017} is based on the phase shifting due to the passage of a gravitational wave from the two laser beams at perpendicular directions. The other effort for detection of gravitational wave is the method of measuring the pulsar timing where using the pulsar timing array (PTAs), it would be possible to detect gravitational waves in the frequency range of $10^{-9} < f/Hz<10^{-7}$ \citep{PTA}. The astrometry of stars inside the Milky Way by GAIA also is suggested as another method that can measure gravitational waves by perturbing metric around the Earth \citep{geri}.

Here in this work, we use the idea of phase-shifting effect by the interaction of a gravitational wave (GW) with 
an electromagnetic wave (EW). The concept is similar to what is used in the interferometry experiments; however,  in our study, this interaction happens in the astronomical scales.  Let us assume a light ray is emitted from a source toward the Earth and a gravitational wave close to the line of sight crosses part of the light ray. This will change the metric of space-time along the trajectory of the light ray. One of the features of this effect would be the frequency perturbation of the source star \citep{kaufmann,mashhoon}. The idea of phase shifting of the electromagnetic wave from a distant coherent source through the interaction with a gravitational wave is proposed by \citet{novikov}. Following this idea, \citet{faraoni} and \citet{bis} used the gravitational lensing concept to calculate the light deflection due to the passage of a light ray through the gravitational wave. However, they have shown that the observational feature of this interaction in the geometric optics is small and that is not detectable with the present instruments.

We take into account the wave optics' features of the electromagnetic radiation from a distant source and calculate the phase distortion as a result of the interaction with the gravitational wave. While the gravitational wave in the cosmological distances is weak, a phase shift of the order of one wavelength in the EW can produce a measurable effect for an observer. We are familiar with the phase shifting effect both in the wave-optics physics in the laboratory scales \citep{born} and in the astronomical scales \citep{schinder, rahvar}. In this work, we investigate the interaction of the EW with the GW where the gravitational wave is sourced from the astrophysical binary stars. Then we calculate the observational features of this interaction as well as the probability of detection of this phenomenon through the observation of quasars with the multiple images from the strong lensing. 

In Section (\ref{EMinC}) we introduce the EW equation in a perturbed metric around the Minkowski and the FRW spaces. From the Helmholtz equation for the propagation of the 
EW, we show that the amplitude of the wave at any point from the Kirchhoff integral is similar to the standard optics where the effective refraction index is replaced with the characteristics of the space-time. The result for the amplitude of EW at any point in the FRW and the Minkowski spaces are similar to each other by replacing the comoving time in the Minkowski space with the conformal time in FRW space.  In Section (\ref{FP}) we use the Fermat Principle to calculate the amplitude of the EW on the observer position both from the geometric optics and the wave optics. We derive the phase shifting from the scattering of the EW from the GW sourced by a binary star
 as well as the amplitude of EW on the observer plane. In Section (\ref{obs}), we investigate the characteristics of the observable parameters of this phenomenon as well as the feasibility of observations in the strong lensed quasars. Conclusions are given in Section (\ref{conc}).

\section{Electromagnetic wave propagation in curve space}
\label{EMinC}
In this section, we study the EW equation and the solution of this equation in the 
perturbed spaces with the background of (i) the Minkowski metric and (ii) the FRW metric. We will use the result of this section for the studying the interaction of EW from a distant quasar with the GWs from the astrophysical sources. 

\subsection{EW propagation in the Minkowski space perturbed by the GWs}
Let us take the perturbation of metric around the Minkowski space by $g_{\mu\nu} = \eta_{\mu\nu} + h_{\mu\nu}$ with the signature of metric $(-,+,+,+)$ (and setting $c=1$)  where  $g^{\mu\nu} = \eta^{\mu\nu} - h^{\mu\nu}$. The propagation of EW in the covariant form is given by $F^{\mu\nu}{}_{;\nu} = 0$ where this equation is rewritin in terms of four-vectors of electromagnetic field (i.e. $A^\mu$) \citep{LL,MTW} as follows:
\begin{equation}
A^{\mu}{}_{;}{}^\nu{}_{\nu} - A^{\nu}{}_{;\nu}{}^{\mu} - R^\mu{}_\nu A^\nu = 0. 
\label{WE}
\end{equation} 
Using the Lorentz gauge of $A^\nu{}_{;\nu} = 0$ and the propagation of light 
in an empty space (i.e. $R^\mu{}_\nu = 0$) , the wave equation reduces to 
\begin{equation}
g^{\alpha\nu}A^{\mu}{}_{;\alpha\nu} = 0.
\end{equation}
We obtain the EW equation coupled to the generic perturbation of the metric as follows: 
\begin{eqnarray}
\label{gwe}
&&\square A^\mu - h^{\nu\alpha}A^\mu{}_{,\nu\alpha} +(h^{\alpha\mu}{}_{,\sigma} + h_\sigma{}^\mu{}_{,}^\alpha - h^\alpha{}_{\sigma,\mu})A^\sigma{}_{,\alpha} \nonumber \\ 
&+&\frac12(\square h_\sigma{}^\mu - h_{\sigma\nu,}{}^{\nu\mu} + \frac12 h^{\mu\nu}{}_{,\sigma\nu})A^\sigma \nonumber \\
&-& (h^{\alpha\sigma}{}_{,\alpha} + \frac12 h^{\alpha}{}_{\alpha,}{}^\sigma)A^\mu{}_{,\sigma}
= 0. 
\end{eqnarray}
For the case that the perturbation of metric results from the gravitational wave, in the transverse-traceless (TT) gauge of $h^\mu{}_\mu = 0$ and  $h^{\mu\nu}{}_{,\nu} = 0$ which satisfies the gravitational wave equation of $\square h^{\mu\nu} = 0$, equation (\ref{gwe})  simplifies to
\begin{equation}
\square A^\mu - h^{\nu\alpha}A^\mu{}_{,\nu\alpha} +(h^{\alpha\mu}{}_{,\sigma} + h_\sigma{}^\mu{}_{,}^\alpha - h^\alpha{}_{\sigma,}{}^\mu)A^\sigma{}_{,\alpha} = 0, 
\label{w2}
\end{equation}
which is identical to the results of \citet{novikov}. We can further simplify this equation for the case that the GW wavelength, $\lambda_g$, is larger than EW wavelength, $\lambda_e$ (i.e. $\lambda_g\gg \lambda_e$). In this case we neglect the derivatives with respect to the perturbations of metric compare to the derivatives with respect to the four-vector of electromagnetic field. Then equation (\ref{w2}) simplifies to 
\begin{equation}
\square A^\mu - h^{\nu\alpha}A^\mu{}_{,\nu\alpha} = 0.
\label{master}
\end{equation}
We note that the $\square$ operator is defined in the Minkowski space and the second term is the perturbation term of this differential equation. 

Let us take $A_0^\mu ={\cal A}_0 e^\mu e^{i\psi_e}$ as the equation for the propagation of EW in the Minkowski space where ${\cal A}_0$ is the maximum amplitude of wave at each point, $\psi_e$ is the phase of EW, and $e^\mu$ is a unit vector along $A^\mu$. The Lorentz gauge results in $e^\mu\partial_\mu\psi_e = e^\mu k^{(e)}_\mu = 0$. For the gravitational wave also, the solution of wave equation in the background of the Minkowski space-time is 
$h^{\mu\nu} = h \epsilon^{\mu\nu}e^{i\psi_g}$  where $h$ is the maximum amplitude of GW at a given point and $\epsilon^{\mu\nu}$ is the polarization of wave. From the Lorentz gauge condition, $\epsilon^{\mu\nu}\partial_\nu\psi_g = \epsilon^{\mu\nu}k^{(g)}_\nu = 0$ and $\epsilon^\mu{}_\mu = 0$. On the other hand, since $h^{\mu\nu}$ satisfies $\square h^{\mu\nu} =0$ , in terms of the polarization tensor and wavenumber can be written as $\epsilon^{\mu\nu}k_\mu^{(g)} k_\nu^{(g)} = 0$. We substitute the background solution of EW equation in the perturbed term of equation (\ref{master}). The overall wave equation simplifies to the Helmholtz equation,
\begin{equation}
\nabla^2A^\mu + ( \omega_e^2 + h\epsilon^{ij}k_i^{(e)}k_j^{(e)} e^{i\psi_g}) A^\mu= 0,
\label{helm}
\end{equation}
where $\omega_e$ represents the frequency of EW in the Minkowski space and the latin indices represent the spatial coordinates of the polarization tensor of the GW.
Now, we define an effective angular frequency for EW in the Helmholtz equation by 
\begin{equation}
\Omega^2 = {\omega_e^2 + h\epsilon^{ij}k_i^{(e)}k_j^{(e)}e^{i\psi_g}},
\end{equation}
where the second term at the right hand side of this equation is smaller than the first term. We simplify the effective frequency of EW to 
\begin{equation}
\Omega = \omega_e( 1+ \frac12 h^{ij}n_i^{(e)}n_j^{(e)}),
\label{omeg}
\end{equation}
where $n_i^{(e)} = k_i^{(e)}/\omega_e$ is the unit vector along the trajectory of the EW  propagation. The solution of Helmholtz equation of $\nabla^2A^\mu + \Omega^2 A^\mu= 0$ with the effective frequency in equation (\ref{omeg}) is give by the Kirchhoff integral, 
\begin{equation}
A^\mu(r) = \frac{1}{4\pi}\int_S\left[A^\mu\frac{\partial}{\partial\hat{n}} (\frac{e^{i\Omega s}}{s})  - \frac{e^{i\Omega s}}{s}\frac{\partial A^\mu}{\partial\hat{n}}\right] dS,
\label{krish1}
\end{equation} 
where the integration is done over the boundary condition, where EW interacts with the GW and for a long distance from the boundary we can simplify this integral to Huygens-Fresnel integral. We note that the effective frequency in equation (\ref{krish1}) may change along the trajectory of the light. We define the overall phase along the trajectory of light as 
\begin{equation}
\int \Omega ds = \omega_e \int (1+ \frac{1}{2} h^{ij}n_i^{(e)}n_j^{(e)})d\ell, 
\label{f1}
\end{equation} 
where $\ell$ follows the trajectory of EW from the source to the observer and the interaction of the GW with EW 
results in a phase shift on the light paths from the boundary to the observer. The integral of  equation (\ref{f1}) is also called the Fermat potential and we will use this function to calculate the observable features both in the wave optics limit and the geometric optics limit (when the wavelength of the EW is small, $\lambda_e \rightarrow 0$).

\subsection{EW propagation in the FRW space perturbed by the GWs}
In this part, we repeat calculation for the FRW space similar to the Minkowski space. For simplicity,  we take a flat spatial curvature for FRW metric (i.e. $k=0$) and conformal with the Minkowski metric \citep{Mukhanov} as $g_{\mu\nu} = a^2(\eta_{\mu\nu} + h_{\mu\nu})$ where $"a"$ is the scalefactor and $g^{\mu\nu} = a^{-2}(\eta^{\mu\nu} - h^{\mu\nu})$ . The electromagnetic equation from equation (\ref{WE}) with the Lorentz gauge and non-empty Universe is given by 
\begin{equation}
A^{\mu}{}_{;}{}^\nu{}_{\nu} - R^\mu{}_\nu A^\nu = 0, 
\label{WE2}
\end{equation}
where at low-redshift Universe and for the case that the wavelength of the GW as well as the wavelength of the EW is smaller than the horizon size (i.e. $k_g/H\gg 1$ and $k_e/H\gg 1$),  we can ignore the second term in equation (\ref{WE2}). Moreover, we assume that the GW wavelength is much larger than the EW wavelength  (i.e. $k_e\gg k_g$). Then equation (\ref{WE2}) simplifies to 
\begin{equation}
\square A^\mu  - 2\frac{a'}{a} {A'}^{\mu} - h^{\nu\alpha}A^\mu{}_{,\nu\alpha}= 0,
\label{FRWe}
\end{equation}
 where $'\equiv \frac{d}{d\eta}$ represents the derivative with respect to the conformal time ($d\eta = dt/a$). Similar to the wave equation in the Minkowski space, the third term at the left hand side of equation (\ref{FRWe}) results from the perturbation of metric. In order to solve this equation, first we write the EW equation in the background of FRW metric, 
 \begin{equation}
 \square A_0^\mu  - 2\frac{a'}{a} {A'_0}^{\mu} = 0,
 \label{emb}
 \end{equation}
 where $A_0^\mu$ represents electromagnetic field in the background of FRW metric. This equation 
 can be solved by defining the new field of $B_0^\mu = a A_0^\mu$. Then equation (\ref{emb}) simplifies to  
 \begin{equation}
B_0''^\mu + (k_e^2- \frac{a''}{a})B_0^\mu = 0,
 \end{equation} 
 where taking into account that $k_e/H\gg 1$, the four-vector of EW has the background solution of $A_0^\mu = {\cal A}_0 e^\mu a^{-1}  e^{i\phi_e}$. Also the gravitational wave has a similar differential equation as (\ref{emb}) \citep{Mukhanov} with the following solution:  
 \begin{equation}
 h^{ij} = h \epsilon^{ij} a^{-1} e^{i\psi_g}.
 \end{equation}
 Substituting the gravitational wave solution in equation (\ref{FRWe}) and using the new field of $B^\mu = a A^{\mu}$, equation (\ref{FRWe}) simples to 
 \begin{equation}
\nabla^2 B^\mu  +\Omega^2 B^\mu = 0. 
\label{HF}
\end{equation}
 where the effective frequency is
 \begin{equation}
 \Omega = \omega_e ( 1 + \frac{1}{2} ha^{-1} \epsilon^{ij} n_i n_je^{i\psi_g} +\frac{1}{2{\omega_e}^2} \frac{a''}{a}),
 \end{equation}
here the second term in the parenthesis is of the order of amplitude of gravitational wave and the third term is of the order of $(\lambda_e H)^2$ where for an EW in the order of one meter wavelength this term is about $10^{-51}$. We can ignore this term compare to the first order perturbation term (second term). The solution of equation (\ref{HF}), substituting the four-vector of electromagnetic field, is 
 \begin{equation}
A^\mu(r) = \frac{1}{4\pi(1+z_g)}\int_S\left[A^\mu\frac{\partial}{\partial\hat{n}} (\frac{ie^{\tilde{\phi}(s)}}{s})  - \frac{ie^{{\tilde{\phi}(s)}}}{s}\frac{\partial A^\mu}{\partial\hat{n}}\right] dS,
\label{krish}
\end{equation} 
where $z_g$ is the redshift of screen at the position of the GW source, and $\tilde{\phi}$ is the overall phase of wave that is sourced from the GW-EW interaction on 
the screen, and $A^\mu(r)$ is the amplitude of the EW at the position of the observer. Again, we define the overall phase of a trajectory of light ray in FRW metric as
\begin{equation}
\tilde{\phi} = \omega_e \int  ( 1 + \frac{1}{2} ha^{-1} \epsilon^{ij} n_i n_je^{i\psi_g}) d\eta.
\label{f2}
\end{equation}
Comparing this equation with the phase in equation (\ref{f1}), the comoving time in the 
Minkowski space is replaced with the conformal time for the FRW metric. Also the 
amplitude of the gravitational wave is scaled with $1/a$ with the expansion of the Universe. Here also we define $\tilde{\phi}$ as the Fermat potential. In the next section, we will investigate the Fermat potential in details and calculate this integral for the interaction of EW with GW, sourced by the binary systems.


\section{Fermat Potential and phase shifting by the gravitational waves}
\label{FP}

\begin{figure}
	\centering
\vspace{-3cm}
	\includegraphics[width=.55\textwidth, angle=0]{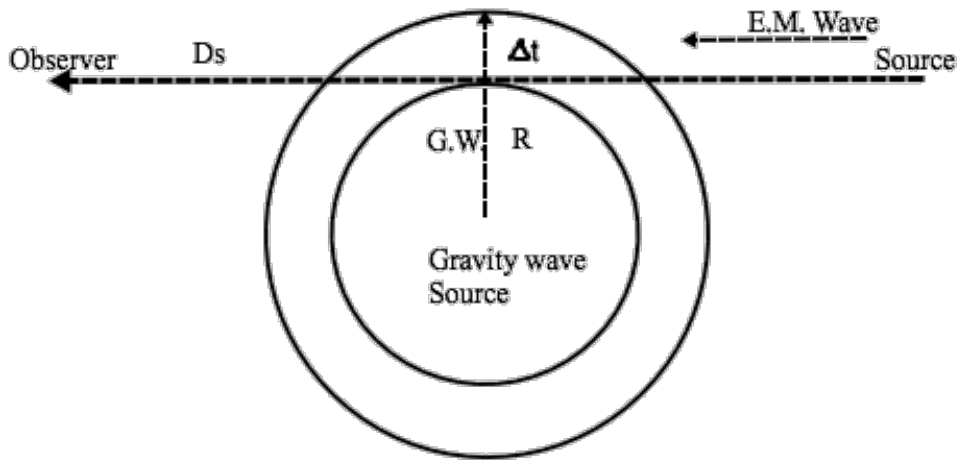}
	\vspace{-4.5cm}
	\caption{Schematic location of a Quasar as the source of electromagnetic wave, observer, and the source of gravitational wave that crosses the light ray along the line of sight of obsever.}
	\label{fig1}
\end{figure}

We start with studying the interaction of GW with a patch of trajectory of  light ray from a source to the observer while it is traveling on a null geodesics. Let us assume a quasar at the distance of $D_s$ from the Earth and an astrophysical source of GW such as a binary star 
produces GW and interacts with the light ray traveling from the quasar to the observer.
Fig. (\ref{fig1}) is the schematic configuration from the quasar, observer and the source of gravitational wave.  The generic form of the perturbation of metric 
around the Minkowski space is written as 
\begin{equation}
ds^2 = (-1+h_{00})dt^2 + 2h_{0i}dtdx^i + (h_{ij}+\delta_{ij})dx^idx^j.
\end{equation}
For the flat FRW metric the formalism is similar to that of Minkowski space, except replacing the comoving time with the conformal time. For the propagation of light ($ds^2=0$), the duration that light ray travels between the source and the observer in terms of overall distance 
is given by 
\begin{equation}
t = \ell + \frac{1}{2}\int (h_{00} + h_{ij}n^in^j + 2h_{0i}n^i)d\ell,
\label{t}
\end{equation}
where $n^i = dx^i/d\ell$ is the component of the unit vector along the direction of the propagation of light.

 We can decompose metric perturbation into the scalar, vector and tensor elements \citep{carrol}. For the condition of the energy-momentum tensor of $\delta T^{i0} = 0$,  perturbation of metric satisfies $h_{00} = h^i_{~i}$. For non-scalar part of metric perturbation, we choose the transverse-traceless gauge of ${{h}_{i0}} = 0$, also ${\hat{h}^{i}{}_{i}} = 0$ where the tensor part of perturbation is assigned with a hat sign. 
Then, the integral for the overall time that light reaches from the source to the observer simplifies to 
\begin{equation}
t = \ell + \int (h_{00} + \frac12 \hat{h}_{ij}n^in^j)d\ell,
\end{equation}
We note that $\ell$ represents the propagation of the light in non-perturbed metric; however it bends locally when their is a local gravity source or gravitational wave crossing the light ray. Here $t$ is proportional to the overall phase of the light traveling from the source to the observer as we discussed in equations (\ref{f1}) and (\ref{f2}).

If we subtract the time duration from that of trajectory of a straight line in the absence of any gravitational field (i.e. $t_0 = \ell_0$), the result of this subtraction (i.e. $\Phi = t-t_0$) is so-called the Fermat potential \citep{schinder} and is given by 
\begin{equation}
\Phi = \frac{1}{2} D |\theta - \beta|^2 + \int (h_{00} + \frac{1}{2} \hat{h}_{ij}n^in^j)d\ell,
\label{phi2}
\end{equation}  
where $h_{00}$ represents the presence of a conventional lens and $\hat{h}_{ij}$ is the contribution of the gravitational wave in phase shifting. Here, $D =D_{d}D_s/D_{ds}$, where $D_{ds}$ is the distance of deflector to the source of EW and $D_d$ is the distance of observer to the deflector, $\beta$ represents the angular position of the source compare to the lens and $\theta$ is the angular position of the images compare to the lens. In general, we may have both a point like lens and a gravitational wave.
The conventional lens might be a galaxy or an isolated star. The contribution of a point-like lens is $h_{00} = 2\phi$ where the potential satisfies the Poisson equation $\nabla^2\phi = 4 \pi G \rho$ and the result of integration for the $h_{00}$ term in the Fermat potential is 
\begin{equation}
\int h_{00} d \ell=\frac{1}{\pi} \int \kappa(\theta') \ln|\theta - \theta'| d\theta'^2,
\end{equation}
where $\kappa(\theta) =\Sigma(\theta)/\Sigma_{crit}$, $\Sigma(\theta)$ is the column density and $\Sigma_{crit}$ is defined as \citep{schinder} $$\Sigma_{crit} = \frac{c^2}{4\pi}\frac{D_s}{D_{ds}D_d}.$$ 
We note that in order to write equations in the FRW metric, we should replace the physical distances in the Minkowski metric with the angular diameter distances in FRW metric. 

\begin{figure}
\centering
	\vspace{-3.4cm}
	\includegraphics[width=.55\textwidth, angle=0]{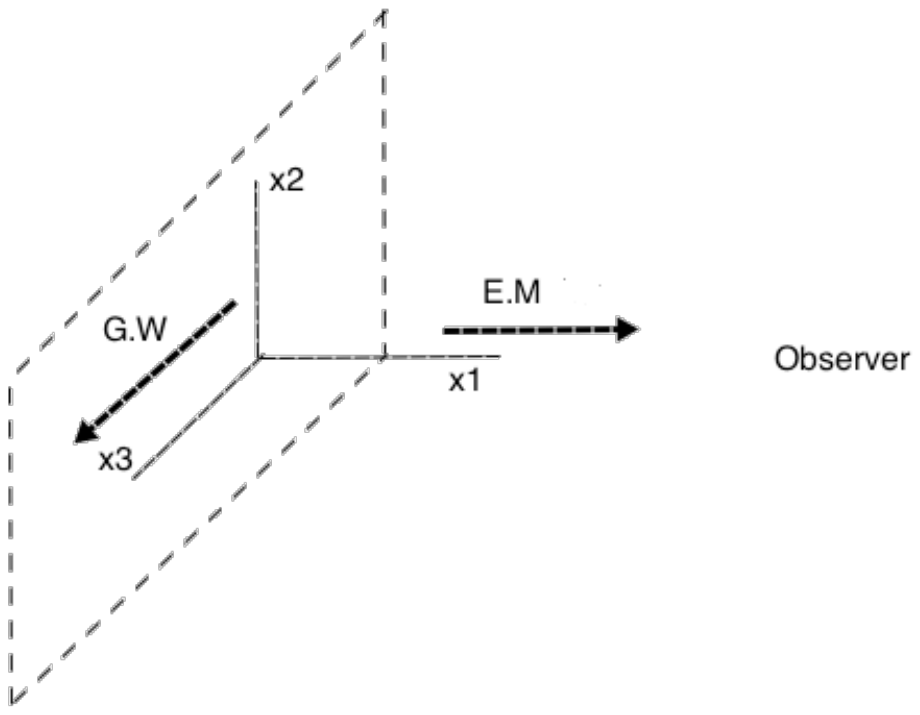}
	\vspace{-4.2cm}
	\caption{Electromagnetic wave propagation in the direction of $x^1$ and the binary system as a source of GW is located on $(x^2,x^3)$ plane, where the orbital plane is at the face-on position with respect to the observer. 
		}
	\label{fig2}
\end{figure}

For calculating the contribution of the gravitational wave in the Fermat potential, we assume the propagation of light in the direction of $x^1$ as depicted in Fig. (\ref{fig2}), where the spatial coordinates are given by $(x^1,x^2,x^3)$ and 
due to the lensing, the direction of light can slightly deviate with respect to $x^1$ axis. 
Let us take the unit vector along the trajectory of the light by $n = (\cos\theta,\sin\theta\cos\varphi,\sin\theta\sin\varphi)$ where $\theta$ and $\phi$ angles are defined in the spherical coordinate and the centre of coordinate is located at the position of the observer. In the absence of gravitational field (in the Minkowski space), the unit vector is $n = (1,0,0)$.

Now we assume a binary system (composed of stars or blackholes) as the source of gravitational wave located on $(x^2,x^3)$ plane. In other word, the orbital plane of the binary is on the face-on position with respect to the observer.  
Substituting $\hat{h}_{ij}$ metric in equation (\ref{phi2}) and assuming a lens with the mass of $M' = M_1 + M_2$ as the total mass of the binary system (i.e. $M_1$ and $M_2$ are the mass of binary components) and the source of GW, the Fermat potential is written as 
\begin{equation}
\Phi= \frac{1}{2}D(\theta - \beta)^2 - 4GM'\ln\theta+\frac{1}{2}\int\left[\hat{h}_{11}(1- 2\theta^2 ) + 2\hat{h}_{12}\theta\right] dx^1.
\label{fermat2}
\end{equation}
From the principle of the least action (i.e. $\delta\Phi/\delta\theta = 0$), the lensing equation is given by  
\begin{equation}
\theta^2 \left(1-\frac{2H_2}{D}\right) - \theta\left(\beta - \frac{H_1}{D}\right) - \theta_E^2 =0, 
\label{mast}
\end{equation}
where $H_1 = \int h_{12}dx^1 \simeq  h_{12} \Delta L$ and $H_2 = \int h_{11} dx^1\simeq h_{11} \Delta L$. Here, $\Delta L$ is the length of the light ray that interacts with the GW and that might be larger than the wavelength of GW (i.e $\Delta L \gg \lambda_g$). In the next section, the numerically calculation confirm our assumption on estimation of $H_1$ and $H_2$. 
 

Equation (\ref{mast}) is the modified conventional lensing equation with single lens where the modification is due to the presence of the GW terms. We can take the GW terms as the perturbation to the standard lens equation. 
The ratio of the perturbation terms to the conventional lensing equation is given by 
\begin{equation} \frac{H}{\theta_E D} \simeq  10^{-16} \left(\frac{n\lambda_g}{1~ly}\right)
\left(\frac{R_s}{3\text{km}}\right)^{-1/2}\left(\frac{D}{1 Gpc}\right)^{-1/2}\left(\frac{\hat{h}_{11}}{10^{-21}}\right),
\end{equation}
where $R_s$ is the Schwarzschild radius of the lens. For the case that the gravity wave results from a binary  system with solar mass stars and one astronomical unit separation, the  wavelength of GW would be about one light year. Assuming that light ray passing at the distance of $1$pc away form the binary system (as the source of GW),  the amplitude of GW is $\hat{h}_{11}\simeq 10^{-21}$, so the perturbation effect of the gravitational wave is negligible in the lensing equation. This means that the astrometric or the light magnification of a source due to the gravitational wave in the lens equation is very small. For the case of the absence of a point mass lens, we set $\theta_E =0$ and from equation (\ref{mast}), the position of image and source relates as $\theta = \beta - H_1/D$. The result is one-dimensional displacement of the position of the source with the amount of $|\Delta \theta| = H_1/D$ with no magnification effect  (i.e. $\mu = 1$) which is in agreement with \citet{bis}. 


In what follows, we perform calculations in the wave-optics regime, where unlike to the geometric optics the gravitational lensing features for the small distortions of wave front is observable. We know this property of wave optics from the interferometry experiments. 
Here, we use the Fermat potential up to the first-order perturbation of the metric and ignore the higher order terms.  Also we ignore the presence of another point mass lens along the line of sight of quasar (i.e. $h_{00} = 0$) as it can be taken as the background field at larger distances from the binary star. Then the Fermat potential simplifies to the following expression:  
\begin{equation}
\Phi(\theta,\beta) = \frac{1}{2}D|\theta - \beta|^2 +\frac{1}{2}\int \hat{h}_{ij}n^in^jd\ell, 
\label{phi}
\end{equation}
where the phase-shifting term along a light ray due to the GW is given by 
\begin{equation}
\delta =  \frac{\pi}{\lambda_e}\int \hat{h}_{ij}n^in^j d\ell,
\label{phase}
\end{equation}
here $\lambda_e$ as the wavelength of EW in denominator is for phase calculation. In what follows, we calculate the amplitude of EW sourced by a binary system.
\subsection{GW produced by the astrophysical binary stars}
Let us assume a binary system with equal masses of $M$ and with the semi-major axis of $a$, located at face-on position with respect to the observer on $(x^2,x^3)$ plane, orbiting around their joint centre of mass with the angular velocity of $\omega_g$. 
The Einstein equation for the gravitational wave with the energy-momentum tensor as the source of GW in TT-gauge of $h^{\mu\nu}{}_{,\nu} = 0$ and $h^{\mu 0} =0$ is given by
\begin{equation}
\Box h^{\mu\nu} = -16 \pi G  T^{\mu\nu}.
\end{equation}
Using the Green function, the tensor components of metric at long enough distance from the source of GW have the following solution: 
\begin{equation}
h^{ij} = \frac{2G}{r}\partial^2_t q^{ij} 
\end{equation}
where
\begin{equation}
q^{ij} = \int T^{00}(x,t-r/c) x^ix^j dx^3,
\end{equation}
is the quadruple moment tensor for the source of GW and for a binary system located at face-on position on $(x^2,x^3)$ plane, that is given by 
\[
q_{ij}=
  \begin{bmatrix}
    0 & 0 & 0 \\
    0 & q_{22} & q_{23} \\
    0         & q_{23}          & -q_{22}
  \end{bmatrix},
\]
where $q_{22} = Ma^2\cos(2\omega_gt_r)$, $q_{23} = Ma^2\sin(2\omega_gt_r)$.  $t_r = t - r$ is the retarded time and $r = \sqrt{(x^1)^2 + b^2}$ is the distance of binary star from the point where the gravitational wave is determined. Here, the impact parameter of EW with the GW is defined by $"b"$ on $(x^2,x^3)$ plane as the closest distance of EW from the source of GW.

 As we noted before, the TT-gauge, (i.e. $h^{ij}{}_{,i} = 0$) implies that $k^{(g)}_ih^{ij} = 0$. So,  the polarization tensor of GW (i.e. $h_{ij}$) should be orthogonal to the direction of the propagation of GW. 
In order to impose this condition, we define the projection tensor of $P^{ij} = \delta^{ij} - \hat{k}^{i(g)}\hat{k}^{j (g)}$ where $\hat{k}^{i (g)}$ is the unit vector along the propagation of GW and $\hat{k}^{(g)} = (\cos\theta',\sin\theta'\cos\phi',\sin\theta'\sin\phi')$. The angles are defined from the position of the source of GW,  where $\theta'$ is the polar angle with respect to the $x^1$ axis and changes in the range of $\theta'\in[0,\pi]$ and $\phi'$ is the azimuthal angle on $(x^2,x^3)$ plane within the range of $\phi'\in[0,2\pi]$. Also, $\theta'$ in terms of the impact parameter is given by $\tan\theta' = b/x^1$. Let us define the transverse-traceless operator \citep{mag} using the projection tensor as
\begin{equation}
P_{jkmn} = P_{jm}P_{kn} - \frac12 P_{jk}P_{mn},
\end{equation}
which guarantees TT-gauge condition of the quadrupole moment as $q_{jk}^{(TT)} = P_{jkmn} q^{mn}$.  Using the projection tensor and the direction of propagation of GW (along $x^1$ axis), the non-zero component of $q_{jk}^{(TT)}$ that enters in the phase shift calculation in equation (\ref{phase}) is 
\begin{equation}
q_{11}^{(TT)} = \frac{1}{4}\sin2\theta'(q_{22}\cos2\phi' + q_{23}\sin2\phi'). 
\label{qq}
\end{equation}



 
 Now we take $n^i = (1,0,0)$ and use $\hat{h}_{ij}^{(TT)}$  to calculate the phase shift of the EW resulting from the perturbation of the metric due to interaction with the GW.  The phase shift is given by 
\begin{equation}
\delta(b,t,\phi') = \frac{\pi}{\lambda_e} \int^{+\infty}_{-\infty}  \hat{h}_{11}^{(TT)}(b,t,\phi',x^1)  dx^1.
\label{psh}
\end{equation}
We note that $t$ is the moment of the interaction of the GW with the EW. We may change this time to the local time measured at the position of the observer by $t_0 = t+D_L-x^1$. Then using $q_{22}$ and $q_{23}$ terms in equation (\ref{qq}), equation (\ref{psh}) can be written as 
\begin{eqnarray}
\label{ph2}
& & \delta(b',t'_0,\phi') = f(M,a,\lambda_e) \times\int_{-D'_{LS}}^{+D'_{L}}\frac{4b'^2\ell'^2}{(b'^2 + \ell'^2)^{5/2}}\nonumber \\ 
&\times& \cos(t'_0 + \ell' -D'_L - \sqrt{b'^2 + \ell'^2} + 2\phi') d\ell', 
\end{eqnarray}
where  $f(M,a,\lambda_e) = 2\pi {R_s^2}/{(\lambda_e a)}$, $R_s$ is the Schwarzschild radius of stars in the binary system with the mass of $M$ and $\ell' = 2\omega_g x^1 $, $t_0' = 2\omega_g t_0$, $b' = 2\omega_g b$,  $D' = 2\omega_g D$ are the new parameters normalized to the $1/(2\omega_g)$.
Also, the numerical value of $f$ is given by 
\begin{equation}
f(M,a,\lambda_e) = 2\pi(\frac{M}{M_\odot})^2(\frac{a}{0.02 a.u.})^{-1}(\frac{\lambda_e}{0.6 cm})^{-1}.
\label{phase4}
\end{equation}
\begin{figure}
	\includegraphics[width=0.5\textwidth, angle=0]{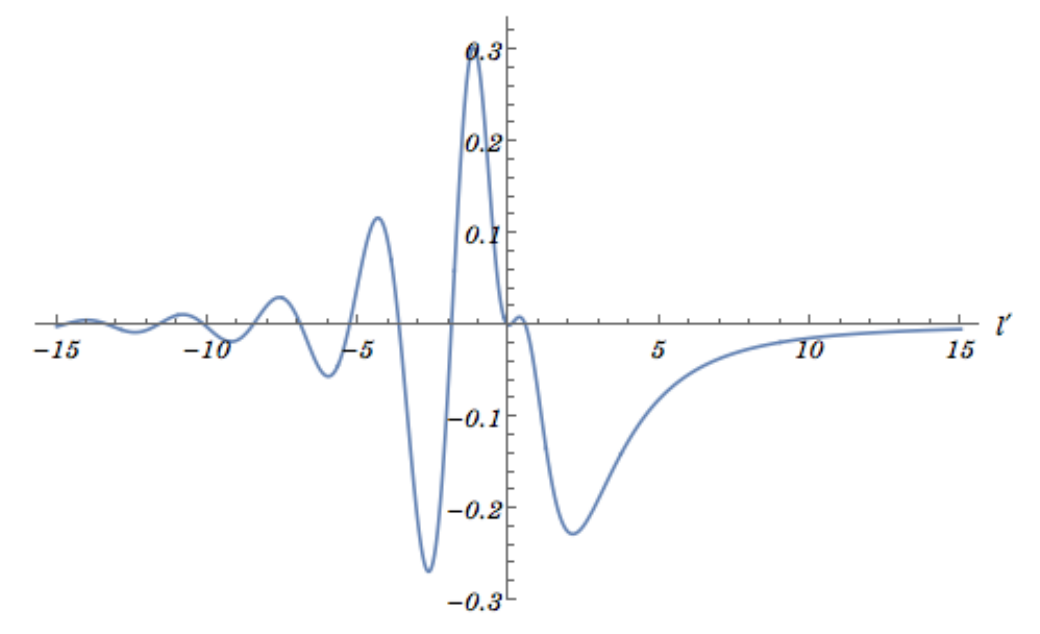}
	\caption{The integrand of equation (\ref{ph2}) as a function $\ell' = 2\omega_g x^1$. Here the normalized impact parameter parameter is set to $b' = 2$ and $\phi'=\pi/3$. } 
	\label{h}
\end{figure}

Before performing the numerical calculation of the phase shifting in equation (\ref{ph2}), we investigate the mathematical behaviour of the integrand in this equation as plotted in Fig. (\ref{h}) (for the parameters of $b' = 2$ and $\phi'=\pi/3$). The physical interpretation of this integrand is that as the EW approaches from the infinity to the source of GW, it interacts with the ripples of the GW with a higher frequency compare to the half of the way when it leaves away from the source of the gravitational wave. So for the domain of $\ell'>0$ the integrand has more contribution in the integral of equation (\ref{ph2}) compare to the domain of $\ell'<0$. Moreover, for the larger distances (i.e. $|\ell'|\gg 1$) the integrand approaches to zero, means that the phase shifting is a local effect within a few wavelengths of GW around the source of GW. We also note that the integrand of  equation (\ref{ph2}) is a function of normalized observer time, $t'_0$ that can change the pattern of the integrand within the time scale of the period of the binary system (i.e. $\Delta t'_0>1$). This means that the phase shifting would be a time dependent pattern within the period of a binary system and for a binary system with the period in the order of year, the observer in a shorter duration of observation will see a static pattern of the phase shifting.

\begin{figure}
\centering
	\includegraphics[width=0.50\textwidth, angle=0]{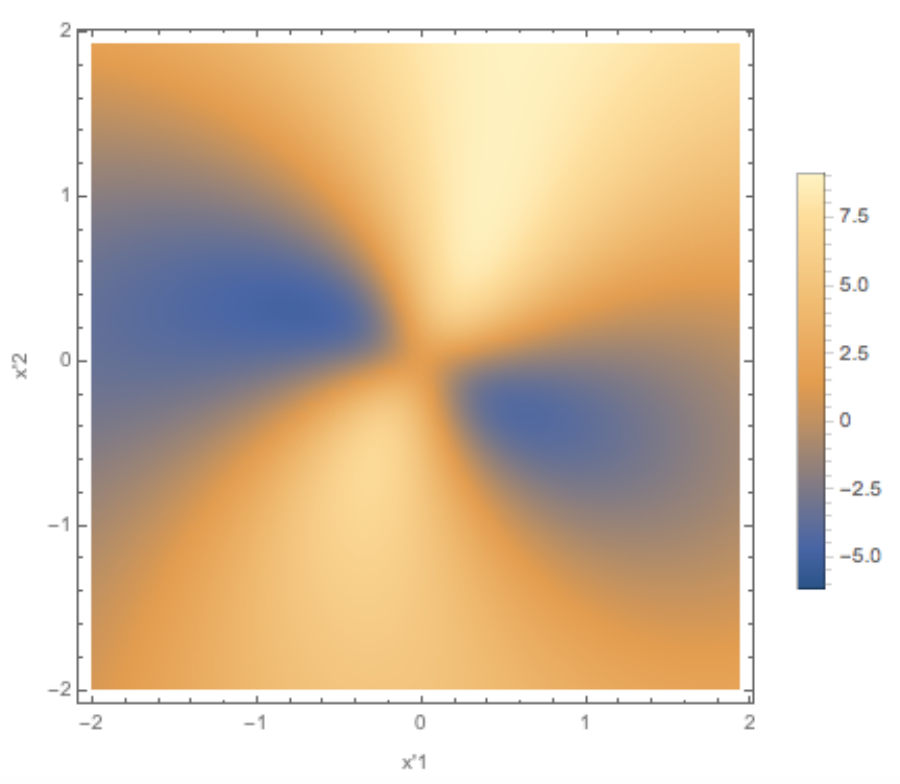}
	\caption{Quadruple phase shifting of the EW wave front on  $(x'^2,x'^3)$ plane, noting that in this space all the physical scales are normalized to $1/(2\omega_g)$. This phase-shifting results from the integration of equation (\ref{ph2}). Here we adapt the following parameters of $t'_0 = 1.5$, $M = 1 M_\odot$, $a = 0.02~$a.u., $\lambda_e = 1~$cm which results in $f = 3.7$ and $\kappa = 0.2$ for generating this phase-shifting pattern.  	
	}
	\label{fringes}
\end{figure}

 Fig. (\ref{fringes}) represents the phase shift on the wave front of EW, resulting from the numerical integration of equation (\ref{ph2}) which has a quadruple structure and we call it as a {\it gravitational grating}. In what follows, we investigate the diffraction pattern from a quadruple gravitational grating on the observer plane. We note that the phase difference on the wave front of EW has to be less than $2\pi$; otherwise, the temporal coherency of the electromagnetic wave for a point-like source will breakdown. From Fig. (\ref{fringes}), we provide a rough estimation on the diffraction patterns on the observer plane. Taking into account that the spatial size of the phase shifting on the gravitational grating is of the order of $\lambda_g$, the scattering angle between the fringes on the observer plane would be in the order of ${\lambda_e}/{\lambda_g}$.



 

In what follows, we investigate the observational effect of the phase shifting of the wave front on the light curve of a distant source, such as a quasar. For a large distance of the observer from the source of GW, the Kirchhoff's diffraction formula in equation (\ref{krish1}) reduces to the Huygens-Fresnel principle  \citep{schinder}  and the magnification of the light on the position of observer can be written as 
\begin{equation}
\mu = \frac{|\int e^{ik_e\Phi(x^2,x^3)} dx^2 dx^3|^2}{|\int e^{ik_e\Phi_0(x^2,x^3)} dx^2 dx^3|^2},
\label{mu}
\end{equation}
where the phase of electromagnetic wave (i.e. $\Phi(x^2,x^3)$) for an arbitrary trajectory of light is given by equation (\ref{phi}) and  $\Phi_0 = \frac{1}{2}D|\theta - \beta|^2$ is the geometrical contribution of the Fermat potential without taking into account the time delay due to the gravitational potential. For simplicity in the integration, we define a characteristic angle for the gravitational wave at distance of $D_{d}$ as $\theta_g = \omega_g^{-1}/2D_{d}$ and replace $\theta$ and $\beta$ 
with $\tilde{\theta} = \theta/\theta_g$ and $\tilde{\beta} = \beta/\theta_g$. Then, equation (\ref{mu}) simplifies to 
\begin{equation}
\mu(\tilde{\beta}_x,\tilde{\beta}_y) =\frac{\kappa^2}{\pi^2}|\int_{-\infty}^{+\infty}\int_{-\infty}^{+\infty}e^{i\kappa|\tilde{\bf\theta}-\tilde{\bf\beta}|^2}e^{i\delta(\tilde{\theta}_x,\tilde{\theta}_y)}d\tilde{\theta}_x d\tilde{\theta}_y|^2,
\label{mu2}
\end{equation}
where $|\tilde{\bf\theta}-\tilde{\bf\beta}|^2 = (\tilde{\theta}_x - \tilde{\beta}_x)^2 + (\tilde{\theta}_y - \tilde{\beta}_y)^2$ 
and $\kappa = \frac{1}{2}k_eD \theta_g^2$ is a dimensionless parameter. The numerical value of $\theta_g$ in terms of physical parameters is given by 
\begin{equation}
\theta_g = 10~nas \left(\frac{M}{M_\odot}\right)^{-1/2}\left(\frac{a}{0.02~a.u.}\right)^{3/2}\left(\frac{D_{d}}{1~Gpc}\right)^{-1}.
\end{equation}
 and $\kappa$ is given by 
 \begin{equation}
 \kappa = \frac{1-x}{3x} \left(\frac{M}{M_\odot}\right)^{-1}\left(\frac{a}{0.02~a.u.}\right)^{3}\left(\frac{\lambda_e}{0.6cm}\right)^{-1}\left(\frac{D_s}{1Gpc}\right)^{-1},
 \end{equation}
where $x=D_{d}/D_{s}$.

\begin{figure}
	\includegraphics[width=0.5\textwidth, angle=0]{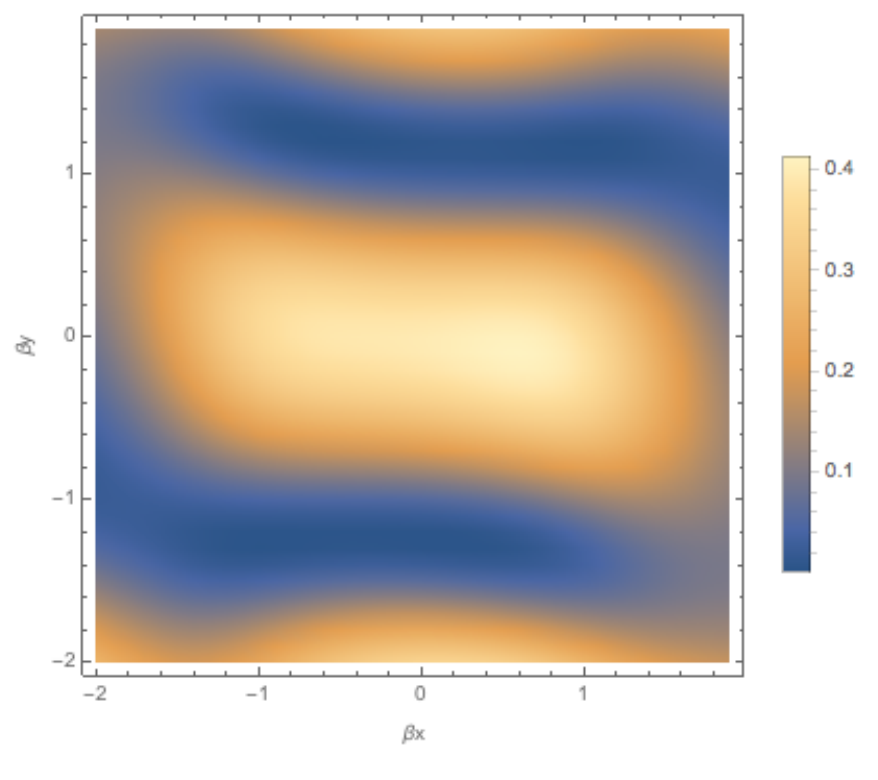}
	\caption{Results from the numerical integration of equation (\ref{mu2}) which represents the interference pattern from a distant quasar on the observer plane. The phase-shifting term in equation (\ref{mu2}) is adapted from figure (\ref{fringes}) with the corresponding parameters. 
	} 
	\label{result2}
\end{figure}

Fig. (\ref{result2}) represents the result of this integration with the phase shifting pattern from the Fig. (\ref{fringes}). We can also write equation (\ref{mu2}) in the polar coordinate system as follows 
\begin{equation}
\mu(\tilde{\beta}) = 4\kappa^2|\int_0^\infty {e^{i(\kappa \tilde{\theta}^2+\delta(\tilde{\theta})) }J_0(2\kappa\tilde{\beta} \tilde{\theta})  \tilde{\theta} d\tilde{\theta} } |^2,  
\label{mu3}
\end{equation}
where the Bessel function in equation (\ref{mu3}) has the wavenumber of $2\kappa\tilde{\theta}$ and since $\tilde{\theta}$ is of the order of unity, the wavelength of ripples in $\tilde{\beta}$ space would be in the order of $\tilde{\beta}\simeq 1/\kappa$. Now, we use the definition of $\tilde{\beta}$ and $\kappa$, the angular separation between the fringes of diffraction pattern on the observer plane would be
$$\Delta\beta \sim \frac{\lambda_e}{\lambda_g}\frac{D_{d}}{D},$$
and since $D_{d}\sim D$, we can simplify the angular separation between the fringes by $\Delta\beta\sim \lambda_e/\lambda_g$.

\section{observational features and Probability of event detection }
\label{obs}
In the first part of this section we discuss about the observational features of the 
GW-EW interaction. 
\subsection{Observational features}
From the observational point of view, we would expect to have an interference pattern on the observer plane  as shown in Fig. (\ref{result2}). Here the EW is sourced from a distant quasar where on the observer plane,  the angular separation between the fringes is of the order of $\Delta\beta\sim \lambda_e/\lambda_g$. Noting that $\beta$ represents the angular position of a quasar with respect to the source of GW, we take into account a dynamics for $\beta$, which represents a relative motion of quasar with respect to the source of GW and the source of EW as follows:
\begin{equation}
\beta(t) = \sqrt{\beta_0^2 + (t-t_0)^2|\frac{{\bf v}_{do}}{D_d} - \frac{{\bf v}_{so}}{D_s}|^2},
\end{equation}
where $\beta_0$ is the minimum impact parameter, $t_0$ is the time of the closes approach of the line of sight to the source of gravity wave, and ${\bf v}_{do}$ and ${\bf v}_{so}$ (according to the convention in the gravitational lensing) are the relative velocities of the source of GW and source of EW with respect to the observer, respectively. Now we define a time-scale that observer crosses the fringes on the observer plane by 
\begin{equation}
\Delta t_f = \Delta\beta \left(\frac{v_{do}}{D_d} - \frac{v_{so}}{D_s}\right)^{-1}.
\label{deltabeta}
\end{equation}
By replacing $\Delta\beta = \lambda_e/\lambda_g$, we can relate the wavelength of 
GW to the wavelength of EW, transit time-scale of fringe crossing, and the relative velocities as follows:
\begin{eqnarray}
\lambda_g &=& 1.6\times 10^{-2} \text{pc} \left(\frac{\lambda_e}{0.6cm}\right)\left(\frac{\Delta t_f }{1h}\right)^{-1} \nonumber \\ 
&\times&\left(\frac{v_{do}}{1000km/s}\frac{Gpc}{D_d} - \frac{v_{so}}{1000km/s}\frac{Gpc}{D_s}\right)^{-1}.
\label{dt2}
\end{eqnarray}
Replacing $\lambda_g = \pi/\omega_g$, where $\omega_g$ is the angular velocity of the binary system as
\begin{equation}
\omega_g = \frac{2\pi}{1~yr}\left(\frac{a}{1a.u.}\right)^{-3/2}\left(\frac{M}{M_\odot}\right)^{1/2},
\label{frequency}
\end{equation} 
and substituting in equation (\ref{deltabeta}) results in transit time as follows
\begin{eqnarray}
\Delta t_f &=& 5.9 h~\left(\frac{\lambda_e}{0.6cm}\right)\left(\frac{M}{M_\odot}\right)^{1/2}\left(\frac{a}{0.02~a.u.}\right)^{-3/2} \nonumber \\
&\times& \left(\frac{v_{do}}{1000km/s}\frac{Gpc}{D_d} - \frac{v_{so}}{1000km/s}\frac{Gpc}{D_s}\right)^{-1}.
\label{dt}
\end{eqnarray}

For a set of typical parameters that we used until now (i.e. $\lambda_e = 0.6$ cm and $\lambda_g \sim 2 \times10^{13}$m) the angular separation between the fringes is about $\Delta\beta \sim  10^{-16}$ rad, and using the transverse velocity of the observer, 
the binary system, and the quasar from their peculiar velocities, the relative angular velocity is of the order of $v_{do}/D_d  - v_{so}/D_s \simeq 1000$ km~s$^{-1}$Gpc$^{-1}$. We use the numerical values for this set of parameters in equation (\ref{dt}) and obtain the modulation in the light curve of quasar in the order of 
$\Delta t_f \simeq 6~$hr.

From the observational point of view there is a limit on detection of time variations of the light curve in which the cadence between the data points should not be smaller than the typical variations of the light curve. Here, we assume the cadence between the data points to be larger than $10$ min (i.e. $\Delta t_f>10$ min). This condition implies a maximum wavelength for detection of the GWs. For instance, using $\lambda_e = 0.6$ cm, from 
equation (\ref{dt2}), and the cadence $>10$ min, results in the condition of $\lambda_g<1.6 \times 10^{-2} pc$. 

\begin{figure}
	\includegraphics[width=0.5\textwidth, angle=0]{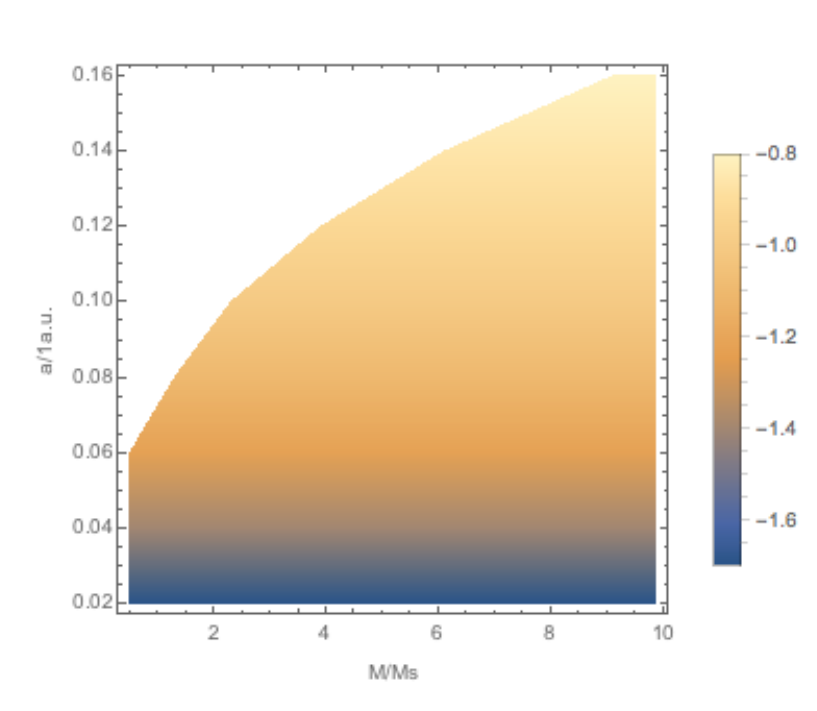}
	\caption{ The variance of the magnification in logarithmic scale (i.e. $\log(\Delta\mu)$) in terms of the semi-major axis of binary system (i.e. $a$) and the mass of binary system (i.e. $M$) where the coloured area represents the coherency condition of the light receiving from the plane of the {\it Gravitational Grating} is satisfied. We adapt the following parameter of $\lambda_e = 6$cm, $D_s = 1$Gpc and $D_d=0.5$ Gpc.} 
	\label{constrain}
\end{figure}

We can also investigate the coherency condition for producing the interference pattern 
from equation (\ref{mu2}) in which the phase shifting of $\Phi = \kappa|\tilde{\bf\theta}-\tilde{\bf\beta}|^2+\delta(\tilde{\theta}_x,\tilde{\theta}_y)$ on the plane of gravitational grating  has to be less than $2\pi$.
 In order to investigate the coherency condition in terms of the parameter space of the binary system, we adapt $\lambda_e = 6$ cm for the wavelength of the observation and assume $a$ and $M$ in the binary system as the free parameters. We also assume the distance of  quasar at $D_s = 1$Gpc and the source of GW at  $D_d=0.5$ Gpc. Fig. (\ref{constrain}) represents the variance of the magnification in logarithmic scale  (i.e. $\log(\Delta\mu)$), in terms of $(a,M)$ parameter space where the coherency condition of EW wave-front is satisfied in the coloured area of this space. The qualitative result from this figure is that, detection of {\it Gravitational Grating} is in favour of the massive binary systems with the larger semi-major axis.


In practice, for the observation of the light curve modulation of a distant quasar from the GW-EW interaction, we propose long-term photometry of the strong-lensed quasars. In the 
strong-lensing systems with the multiple images from a quasar, by time shifting the light curves of images with the amount of time delay, we can remove the intrinsic variations of a quasar to extract our desired foreground signals. This method has been used for detection of the microlensing signals \citep{qs} as well as the weak-lensing effect by the dark micro-halos \citep{rahvar2} in the quasars light curve. Since the time-scale of microlensing and dark micro-halo transits are longer than the transit time-scale of diffraction fringes from the GW, in principle we can filter out these two backgrounds.

\subsection{ Probability of detection of GW-EW interaction for a given quasar at Gpc distance}


We have seen that the diffraction angle of the electromagnetic waves from the gravitational waves is of the order of $\Delta\beta \simeq \lambda_e/\lambda_g$. Comparing this diffraction angle with that in the double slit Young experiment, with the separation of $"d"$ between the slits indicates that the wavelength of gravitational wave (i.e. $\lambda_g$) plays the role of the silts (i.e. $d$) in the Young experiment. That is why we called this phenomenon as the {\it gravitational grating}.  The minimum size of the grating plane is of the order of $\lambda_g$ and the cross-section associated to this interaction is given by $\sigma\simeq \pi \lambda_g^2$. 
 Now, similar to the gravitational microlensing \citep{pac,rahvar2015}, we define the optical depth as the probability of crossing the line of sight with the cross section of the {\it gravitational grating}. The main difference between the definition of the optical depth in the microlensing with that of gravitational grating is that the optical depth for the microlensing is independent of the mass of the lenses and it depends only to the average mass density of lenses, however in our case, the optical depth depends on the wavelength of the GWs. Let us assume $f(\lambda_g)$ as the distribution function of the binary systems that produces GW with the wavelength of $\lambda_g$. We define the optical depth  as follows:
\begin{equation}
\tau(\lambda_e) = \frac{\int_{0}^{D_s} \int_{\lambda_g^{min}}^{\lambda_g^{max}} \sigma(\lambda_g) n_{bs}(x^1) f(\lambda_g) d\lambda_g  dx^1
}{\int_{\lambda_{g,0}^{min}}^{\lambda_{g,0}^{max}} f(\lambda_g) d\lambda_g},
\label{tau}
\end{equation}
where $n_{bs}(x^1)$ is the number density of binary stars along the line of sight 
 at the position of $x^1$ that produces GW. Also, $\lambda_g^{min}$ and $\lambda_g^{max}$ are the minimum and maximum wavelengths of the GW where interaction with the EW results in the wave optics features of the EW. The range for $\lambda_g$ depends on the EW wavelength that we are using for the observation. Also $\lambda_{g,0}^{min}$ and $\lambda_{g,0}^{max}$ are the generic possible range of the wavelengths for the GW that binary stars can produce. We note that $\lambda_g$ for a binary star is a function of the mass and the semi-major axis of the companions. 


In order to apply $f(\lambda_g)$ in equation (\ref{tau}), we simplify our calculation by assuming that the two companions of the binary systems have equal masses (i.e. $M_1/M_2 = 1$) and zero orbital eccentricity. The distribution function for the semi-major axis of the binary stars (i.e. $a$) is given by the  {\"O}piks law \citep{opik} as follows
\begin{equation}
f(a) \propto a^{\gamma}~~~~ \text{where}~~~~a_{min}<a<a_{max}.
\end{equation}
We can rewrite this equation in terms of the mass of companions and the orbital period of the binary system as 
\begin{equation}
f(P,M)\propto M^{(\gamma + 1)/3}P^{(2\gamma - 1)/3}.
\end{equation}
Here we adapt the conventional value of $\gamma = -1$ \citep{opik2}, which results in a mass-independent distribution function for the orbital period of binary stars as $f(P) \propto P^{-1}$ or in terms of the GW wavelength $f(\lambda_g)   \propto \lambda_g^{-1}$. Substituting this distribution function in equation (\ref{tau}) and using a none-zero value for the number density of binary systems in the strong lensing galaxy, the optical depth obtain as follows: 
\begin{equation}
\tau(\lambda_e) = \frac{\pi n_0 w}{2} \frac{(\lambda_g^{max})^2 - (\lambda_g^{min})^2}{\log(\lambda_{g,0}^{max}/{\lambda_{g,0}^{min}})}, 
\label{tau3}
\end{equation}
where $\omega$ is the size of strong lensing galaxy that the trajectory of light of one of the images crosses it and $n_0$ is the average number density of binary stars in a typical galaxy. We adapt the possible range for the period of the binary stars \citep{opik2} from $P_{min}^0 = 0.5$ d to $P_{max}^0 = 0.15$Myr. Substituting the numerical range of GW wavelength both for $\lambda_g$ and $\lambda_{g,0}$, for the EW wavelength of $\lambda_e = 0.6$ cm, equation (\ref{tau3}) results in 
\begin{equation} 
\tau \simeq 0.2 \left(\frac{n_0}{1pc^{-3}}\right)\left(\frac{w}{10kpc}\right). 
\end{equation}
Taking $n_0 \simeq1 pc^{-3}$, as the typical number density of stars inside a galaxy where more than half of them are binaries, the result of optical depth calculation means that having a quasar with the multiple images due to the strong lensing by a galaxy with the size of $10$~kpc , the probability for the detection of GW-EW interaction is about $20$ percent.

\section{conclusion}
\label{conc}
In this work, we proposed an indirect observational method for detection of the gravitational waves (produced by a binary star) via the interaction with the electromagnetic wave from a distant source such as a quasar. We have investigated the 
propagation of the electromagnetic waves in the perturbed Minkowski and FRW spaces and derived a generic differential equation where for the specific conditions, we recovered the results from the previous studies. The formalism in these two spaces are similar, expect that we replaced the comoving time in the Minkowski space with the conformal time in the FRW space. Also the amplitude of fields decay with the expansion of the Universe. 

We have shown that the solution of the differential equation for the electromagnetic field is a Kirchhoff integral where the phase of the field can be replaced with the concept of the Fermat potential. For the case that the source of electromagnetic wave and the source of gravitational wave are located at large enough distances from the observer, the integral simplifies to the Huygens-Fresnel principle. The formalism is similar to the standard optics expect replacing the refraction index in the optics with the perturbations of the metric in our study. Here in the work, we introduced the concept of {\it gravitational grating}, where similar to the optical grating can produce phase-shifting and the diffraction pattern on the observer plane.

For the limit of $\lambda_e \rightarrow 0$, we can recover the results of the geometric optics
and unlike to the wave optics effect, the geometric optics effects is too small to be observed. Our results in this part was consistent with the previous studies.  
Taking into account the wave optics effects, the distortion of the wavefront, in the order of 
one wavelength of the electromagnetic radiation can produce the diffraction pattern with the angular separation  in the order of $\lambda_e/\lambda_g$ between the fringes. We emphasized that while this phase-shifting is essential for detection of the interference pattern, however we should be careful about the temporal coherency where for larger phase-shifting the wave optics effects faded out.

 Taking into account a relative motion of the observer-the source of gravitational wave and a quasar results in that the observer moves through the diffraction pattern on the observer plane. 
 We have shown that the observations in the milli-meter wavelength results in a modulation of the light curve in the order of few hours. The practical procedure for the observation of this phenomenon is the long term monitoring of the multiple images of the quasars in the strong  lensing systems, where some of images cross the lensing galaxy. The advantage of this method is that (i) we can 
 remove the intrinsic variations of the quasars by time-delay shifting and (ii) for those images cross the strong lensing galaxy, the optical depth of the electromagnetic wave-gravitational wave interaction is around $0.2$.  This method can open a new window for indirect detection of gravitational waves.  






I would like to thank Marc Moniez, Shant Baghram, Hessamaddin Arfaei, Viktor Toth and Bahram Mashhoon for their useful comments. Also I would like to thank anonymous referee for his/her useful comments. This work was supported by Sharif University of Technology's Office of Vice President for Research under Grant No. G950214.

\vspace{-0.5cm}

 \bibliographystyle{mnras}

  \bibliography{ref}

\label{lastpage}

\end{document}